\documentclass[twocolumn,superscriptaddress,prl,aps,showpacs,amsmath,amssymb]{revtex4}
\usepackage{graphicx}
\begin{document}

\title{Minimal fragmentation problem}
\author{Fernando Parisio} 
\affiliation{Departamento de F\'{\i}sica, Universidade Federal de Pernambuco, 50670-901,
Recife, Pernambuco, Brazil}
\author{La\' ercio Dias}
\affiliation{ Departamento de Estat\'{\i}stica, Universidade 
Federal de Pernambuco, 50740-540, Recife, Pernambuco, Brazil}


\begin{abstract}
As an alternative to the paradigmatic fragmentation problem of a single object crushed into a great number
of pieces, we survey a large collection of identical bodies, each one randomly split 
into {\it two} fragments only. While some key features of usual fragmentation are preserved, this
minimal approach allows for closed analytical results on both, shape abundances and mass 
distributions for the fragments, with robust power-law regimes. All the results are compared to
numerical simulations.     
\end{abstract}
\pacs{05.40.-a,02.50.-r,46.50.+a}
\maketitle


The generic designation complex system fits
few subjects as well as fragmentation. Though the interest in this
field grew from specific examples \cite{mott}, it now spreads from basic physics 
to material science and geology, including multifragmentation of nuclei \cite{mekjian}, fracture 
propagation in diverse materials \cite{astrom}, and earthquakes \cite{geology}. 
The typical fragmentation problem 
consists in the catastrophic crushing \cite{grady} of a body due to the sudden 
appearance of fracturing stresses, the main goal being to determine the 
mass distribution of the resulting myriads of pieces.  The outcome of such a process strongly
depends on a large number of factors and their mutual interactions. Among them, 
the mechanical properties of the material, e. g., brittle \cite{gladden} or plastic \cite{hermann1}, 
its effective dimensionality \cite{ishii-linna}, the mechanism of fragmentation 
(explosion, impact by a projectile, uniform compression, etc), and the magnitude of the 
energy input \cite{ching}. This multitude of variables generates a rich pool of data
which has often been observed to follow power laws. 
A fully analytical approach to this class of problems is far beyond our present
capabilities, both, due to mathematical complexity and to the lack of a complete knowledge on 
the physical responses that lead to crack nucleation and propagation under different conditions. 

Although the reduction of an intricate problem to a simpler, but non-trivial one, is a
common {\it modus operandi} of physics, this was not fully carried on in the field of fragmentation
(see however the classical works by Gilvarry \cite{gil}).
Instead of considering one object broken into an enormous number 
of pieces we focus on an set of identical bodies minimally segmented, i. e., 
each one randomly split into two fragments only. 
Suppose you take a tile and throw it from a modest height so that the scenario 
where a single crack develops is the most likely. 
\begin{figure}[ht]
\includegraphics[width=1.3 in, angle=0]{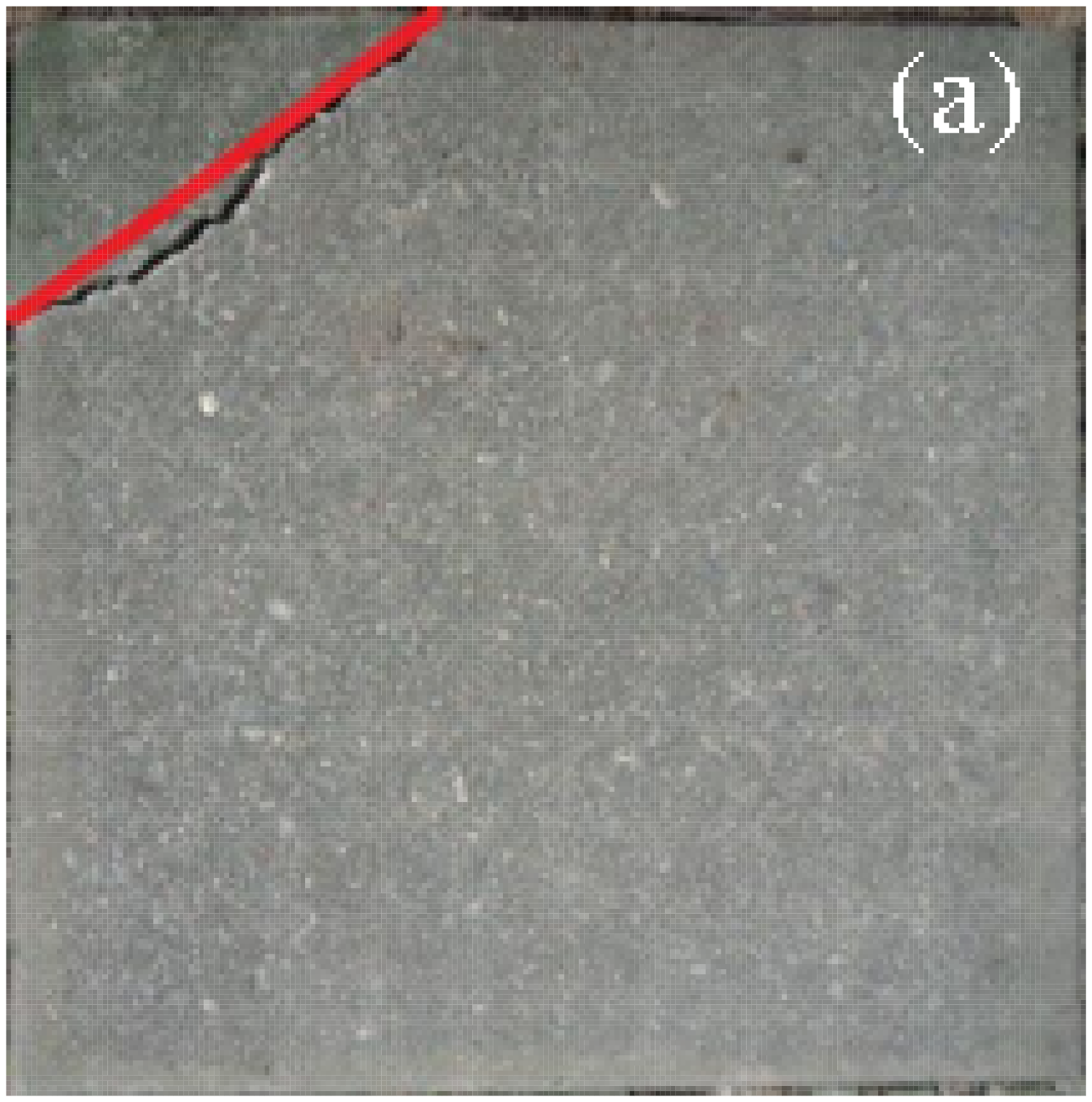}
\includegraphics[width=1.3 in, angle=0]{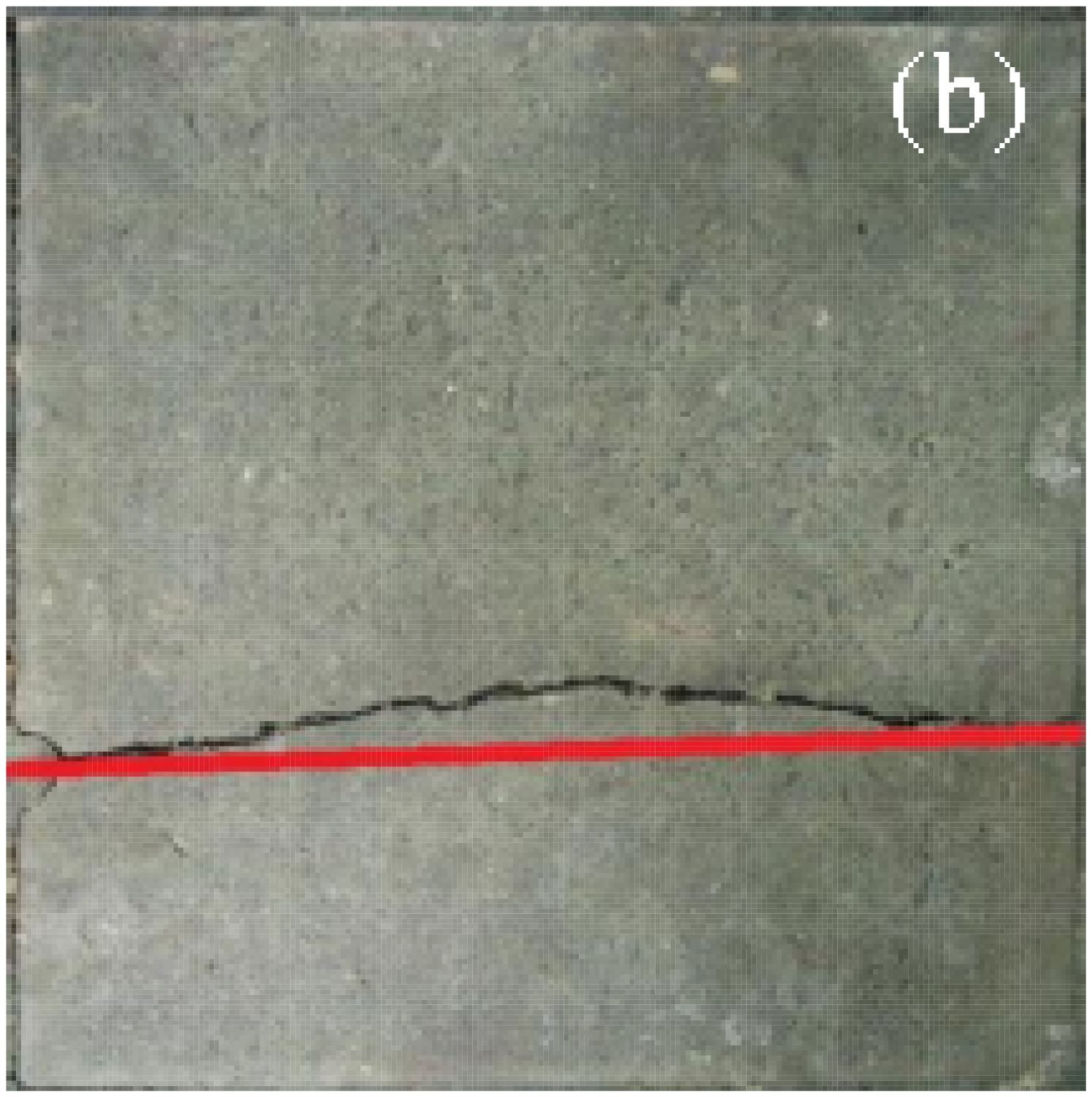}
\includegraphics[width=1.3 in, angle=0]{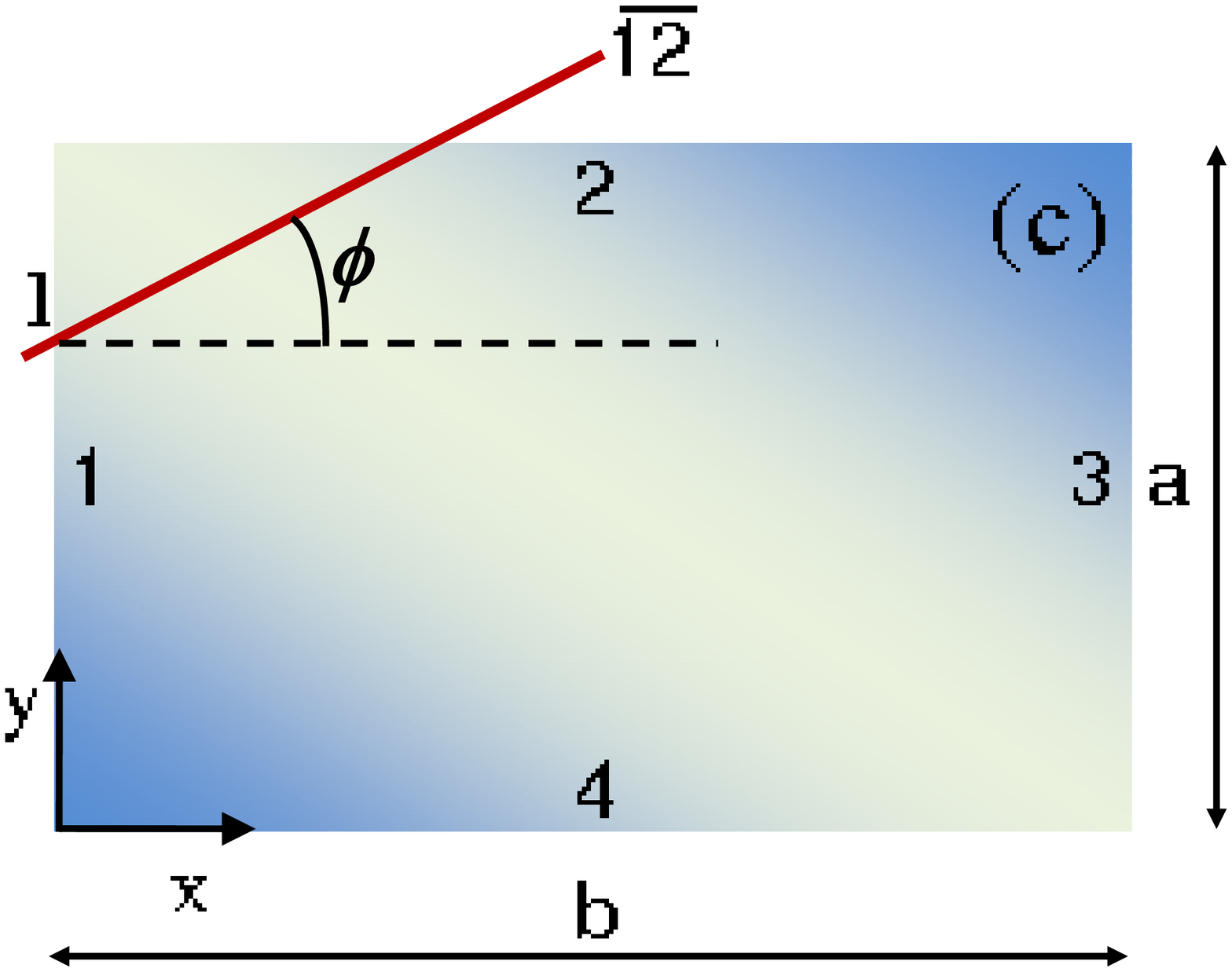}
\caption{(color online) Minimally segmented tiles. In (a) the fragments are a triangle and a pentagon, while in (b) 
two trapezoids are obtained. In (c) we represent a minimal fragmentation. The variables $l$ and $\phi$ are shown along 
with the adopted labeling of the sides.}
\label{fig1}
\end{figure}
By repeating this procedure with a large number of elements and selecting those minimally fragmented, 
one can obtain statistical information on the shape and mass distribution of the resulting shards in 
the same way as it is done in usual fragmentation problems. In fact, the experimental setup presented in Ref. 
\cite{santos} on flat bodies
laterally impacted, seem appropriate to generate this kind of ensemble in a controlled way (for spherical bodies see \cite{experiment}).
A further example of a spontaneous process that gives rise to minimal fragmentation is the propagation 
of a fracture on a tiled floor. 

We will be concerned with 2D objects, namely, rectangles with 
side lengths $a$ and $b$, defining an aspect ratio $\gamma=b/a$. 
We notice that minimal segmentation is naturally associated to a low energy process, 
the cracks are neither branched nor too irregular [see figs. \ref{fig1}(a) and \ref{fig1}(b)]. 
For this reason, in our geometric statistical model, they are represented by a random segment for each elementary rectangle. 
We must make the meaning of these segments completely unambiguous. As discussed in several 
works, there is no {\it a priori} geometric statistical prescription to define the partitioning of a body
with dimension greater than one. Different rules lead to quite distinct results, 
a fact that is strikingly illustrated by the Bertrand Paradox \cite{kendal}. 
Since we are interested in situations where the cracks either originate or cross the boundaries of 
rectangles without any bias and isotropically, our basic assumptions are: the segmentation line has 
a uniform probability to cross any interval in the boundary of the rectangle, the angular 
distribution of the segments being also uniform. Therefore, in order to generate a line we define two
random variables $l$ and $\phi$ uniformly distributed in the intervals $[0,2a+2b] \equiv [0,L]$ 
and $[-\pi/2,\pi/2]$, respectively. For definiteness, $\phi$ is given as the angle between the segment
and the normal direction to the side selected by the variable $l$ [see Fig. \ref{fig1}(c)]. 
This construction encompasses the desired physical inputs and, at the same time, employs the minimum 
possible set of variables. 

\begin{figure}[ht]
\includegraphics[width=1.5 in, angle=0]{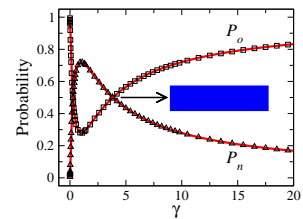}
\caption{(color online) Probabilities of obtaining two trapezoids ($P_o$) and a triangle plus 
a pentagon ($P_n$) as functions of $\gamma$. For $\gamma \approx 3.83$ these probabilities are equal. }
\label{fig2}
\end{figure}

It has been recently suggested that geometric properties, in opposition to mass 
distributions, may be a more stringent test to theoretical models \cite{santos}. 
In this regard the shape of the resulting fragments can be classified according to 
the way a single line can split a rectangle. The two alternatives are: 
(i) the intersection points with the border are on 
neighboring sides or (ii) the segment cross opposite sides. 
Within our approximation, scenario (i) produces 
as fragments a triangle plus a pentagon, while situation (ii) leads to two trapezoids, all irregular 
in general. We begin by addressing the following question: 
what is the relative frequency of events (i) and (ii) as a function of the aspect ratio $\gamma$
for a large number of events? 
We denote all (statistically) relevant crossings as follows: the segment $\overline{jk}$
intersects sides $j$ and $k$, with $j, k =1,2,3,4$, [Fig. \ref{fig1}(c)] the associated
probabilities given by $P_{jk}(\gamma)$.
Scenario (i) occurs if the segment is of the type $\overline{13}$ or  $\overline{24}$.
For situation (ii) the possibilities are $\overline{12}$, $\overline{23}$, $\overline{34}$, 
$\overline{41}$. Thus, the probability of obtaining a crack that connects neighboring sides is
mapped into
$P_n=P_{12}+P_{23}+P_{34}+P_{41}$. From the symmetry of the problem it is evident that these 
probabilities are equal, so
$P_n=4 P_{12}$. To calculate $P_{12}$ we note that $\overline{12}$ may arise in two ways.
First, the point drawn is on side $1$ ($l \in [0,a]$) and the angle satisfy $\cot
^{-1}[b/(a-l)] < \phi <\pi/2 $, which gives the angular interval $\Delta \phi =\pi/2-\cot
^{-1}[b/(a-l)]=\tan^{-1}[b/(a-l)]$. Since the angular distribution is uniform, 
the probability of obtaining $\overline{12}$ in this first case is 
\begin{eqnarray}
\nonumber
F(\gamma)=\frac{1}{\pi L}\int_{0}^{a}  \tan^{-1}\left( \frac{b}{a-l}\right) {\rm d}l\\
=\frac{\gamma}{2\pi(\gamma + 1)}
\left[ \frac{1}{\gamma} \tan^{-1} (\gamma)+\frac{1}{2}\ln\left(1+\frac{1}{\gamma^2} \right) \right]\;.
\label{eq1}
\end{eqnarray}
The second possibility of getting $\overline{12}$ is $l \in [a,a+b]$ and $-\pi/2 < \phi < \cot
^{-1}[a/(l-a)]$, that yields the same result as
in (\ref{eq1}), with $a \leftrightarrow b$, i. e., $\gamma \rightarrow 1/\gamma$, as expected.
We then get $P_{12}=F(\gamma)+F(1/\gamma)$, leading to
\begin{equation}
P_n=4F(\gamma)+4F(1/\gamma)\;.
\label{p12}
\end{equation}
The probability of two opposite sides traversed by the random segment (two trapezoids) is
$P_o=1-P_n$. In Fig. \ref{fig2} we show these probabilities as functions
of $\gamma$. The continuous curves represent $P_n$ and $P_o$. The open triangles and squares correspond to 
numerical experiments with $10^6$ segmentation events (for each point). Note that $P_n$ attains its 
maximum value for $\gamma =1$, where 
\begin{equation}
P_n=\frac{1}{2}+\frac{\ln 2}{\pi} \approx 0.72\;.
\end{equation}
This probability remains larger than $P_o$ for $\gamma< 3.83$, the aspect ratio of the rectangle highlighted in Fig. \ref{fig2}. 

Note, however, that these results say nothing about the size of the fragments. Obtaining an
experimental plot compatible to Fig. \ref{fig2} is not as simple as it may appear. In order to get
the right frequencies, tiny fragments (which can be easily overlooked) must be counted as much as larger ones. 
The point is that the density of very small triangles diverges as the fragment mass approaches zero,
a signature of usual multifragmentation. This might not be expected at a first glance, since our model seems to 
produce mainly sizable fragments. 
This naturally leads us to the central fragmentation problem of mass distributions. 
In experiments it is much easier to record the accumulated probability ${\cal P}_{>}(m)$ for a fragment to have 
a mass greater than $m$. However, for convenience we will use the complementary accumulated 
probability ${\cal P}(m)=1-{\cal P}_{>}(m)$:
\begin{equation}
\label{P-p}
{\cal P}(m)=\int_0^{m}{\rm d}m' \,p(m')\;,
\end{equation}
where $p(m)$ is the associated probability distribution.
Let $2M= ab$ be the total mass of the rectangle (without loss of generality we set the
superficial density of mass to $1$).
Since in each event we have a minimal segmentation, for every fragment of mass $m$ there is 
another one of mass $2M-m$, that is,  ${\cal P}(m)={\cal P}_{>}(2M-m)=1-{\cal P}(2M-m)$.
In particular, for $m=M$ we must have ${\cal P}(M)=0.5$. Given this symmetry, we only need to consider 
values of $m$ between $0$ and $M$, meaning that, in each event, it is sufficient to collect 
the smaller of the two pieces, which is either a triangle or a trapezoid, to get all 
information.

\begin{figure}[ht]
\includegraphics[width=2.7 in, angle=0]{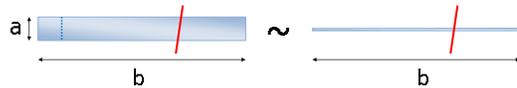}
\caption{(color online) For an elongated body the ratio between the fragment mass and total mass is nearly the same 
as in the $1D$ case. The square delimited by the dashed line gives the minimum area that characterizes 
a ``large" fragment.}
\label{fig3}
\end{figure}

We proceed by defining the auxiliary quantity ${\cal P}(l,m)$, the accumulated probability for a {\it fixed}
value of $l$, i. e.,
\begin{equation}
{\cal P}(m)=\frac{1}{L}\int_0^{L} {\cal P}(l,m) {\rm d} l\;.
\end{equation}
To determine ${\cal P}(l,m)$ we must specify the mass $m'<m \in [0,M]$ of an arbitrary smaller fragment 
in terms of the random variables $l$ and $\phi$. Let us suppose once more that side $1$ is selected, $l \in [0,a]$. 
In this case, if the fragment is a triangle, then $m'=(1/2)(a-l)^2\cot \phi$, while for a trapezoid
we get $m'=b(a-l)-(1/2) b^2 \tan \phi$. Note also that if $m<b(a-l)/2$, only triangles are possible. 
Given $m$ and $l$ the limiting angle for triangular fragments is $\phi_{lim}=\tan^{-1}[(a-l)^2/2m]$.
Again, since the angular distribution is uniform, the probability is proportional to the allowed angular
interval $\Delta \phi=\pi/2 - \phi_{lim}$. Thus, we get
\begin{equation}
{\cal P}(l,m)=\frac{1}{\pi}\left[  \frac{\pi}{2} 
- \tan^{-1}\left( \frac{(a-l)^2}{2m}\right)\right] \;,
\end{equation}
for $0 < m < b(a-l)/2 $, that is, $0 < l < a-2m/b$. 
Accordingly, if $b(a-l)/2< m < ab/2$ we must have a trapezoid, leading to
\begin{equation}
{\cal P}'(l,m)=\frac{1}{\pi}\left[  \frac{\pi}{2} 
+ \tan^{-1}\left( \frac{2m}{b^2}-\frac{2(a-l)}{b}\right)\right] \;.
\end{equation}
From these results we obtain the following accumulated probability with $l$ restricted
to side 1:
\begin{equation}
{\cal P}_1(m)=\frac{1}{L}\int_{0}^{a-\frac{2m}{b}}{\cal P}(l,m) {\rm d} 
l+\frac{1}{L}\int_{a-\frac{2m}{b}}^a{\cal P}'(l,m) {\rm d} l\;.
\end{equation}
The full accumulated probability is then ${\cal P}(m)= {\cal P}_1(m)+{\cal P}_2(m)+
{\cal P}_3(m)+{\cal P}_4(m)= 2{\cal P}_1(m)+2{\cal P}_2(m)=2{\cal P}_1(m)
+2{\cal P}_1^{a \leftrightarrow b}(m)$, where the equalities are justified by direct symmetry
arguments. The final result can be compactly written as
\begin{eqnarray}
\label{acum}
\nonumber
{\cal P}(\mu , \gamma)= 1-\frac{2}{\pi (1+ \gamma)}
\sum_{j,k=0,1}\gamma^{1-j} (-\zeta_j)^{1-k} \tan^{-1}\left(\zeta_j^{1-2k}\right)\\
-\frac{\sqrt{2 \zeta_1}}{\pi (1+ \gamma)}\sum_{j,k,q=0,1}(-1)^{q}\tan^{-1}\left[\sqrt{2\zeta_j^{1-2q}}+(-1)^k\right] \;,
\end{eqnarray}
where $\zeta_0=\mu/\gamma$, $\zeta_1=\mu \gamma$, and $\mu=m/M$ is the dimensionless mass
with $\mu \in [0,1]$. The normalization is such that ${\cal P}(1,\gamma)=1$. Inverting Eq. (\ref{P-p}) and using Eq. (\ref{acum}) we get
the following probability distribution
\begin{eqnarray}
\nonumber
\label{density}
p(\mu , \gamma)=\frac{{\rm d}{\cal P}}{{\rm d}m}= \frac{{\cal P}(\mu, \gamma)-1}{2 \mu}\\+\frac{1}{ \pi \mu (1+ \gamma)}
\sum_{j,k=0,1}\gamma^{1-j} \zeta_j^{1-k} \tan^{-1}\left(\zeta_j^{1-2k}\right) \;.
\end{eqnarray}
Therefore, we obtain all information on the ensemble in a fully analytical way. We find the density $p$ 
to be essentially a composition of two power laws connected by a stretched dimensional crossover.
First, in the limit of very small fragments, which we call the dust regime, we find that $p$ asymptotically diverges as
\begin{equation}
\label{dust}
p \sim \left[ \frac{\sqrt{2 \gamma}}{(1+ \gamma)}\right]\, \mu^{-1/2}\;, \; \mbox{for} \; \mu \rightarrow 0\;,
\end{equation}
showing that the model leads, chiefly, to small fragments.
Interestingly enough, this same exponent has been reported in numerical simulations of a number of models for 
multifragmentation of planar objects \cite{santos0}. This behavior is completely dominated by triangular fragments since the 
probability of getting a trapezoid becomes vanishingly small for $\mu \rightarrow 0 $. 
The final regime is observed for $\gamma \ne 1$, becoming more evident for greater values of $\gamma$. 
For massive fragments we get
\begin{equation}
\label{large}
p \sim 2\left[\frac{\gamma \tan^{-1}\gamma + \cot^{-1}\gamma}{\pi(1+\gamma)}\right]\, \mu^{0}\;, \; \mbox{for} \; \mu \rightarrow 1\;.
\end{equation}
The transition from (\ref{dust}) to (\ref{large}) is due to a change in effective dimensionality. To understand this, consider the minimal fragmentation of a thin, uniform, rod with unit size and mass. Let $l$ be a random variable uniformly distributed on $[0,1]$, the value of $l$ defining the point at which the rod breaks. It is evident that the mass distribution of smaller fragments directly follows the distribution of $l$ itself, i. e., $p(l)\propto l^0 \Rightarrow p(\mu) \propto \mu^0$. The fact is that, especially for higher aspect ratios, 
only small fragments realize the 2D character of the elongated rectangle, for larger fragments the behavior of the probability density is virtually the same as in the 1D case. See Fig. \ref{fig3}, where the ratio $m/M$ is nearly the same in both depicted situations. 
For a fragment to be considered large its characteristic size should overstep the length of the shorter side of the rectangle, i.e., its area must exceed $a^2$ in comparison to the total area $ab$ (or $a^2/2$ in comparison to $M$), which leads to $\mu \sim a^2/ab=1/\gamma$, so, for $\mu > 1/\gamma$, 
$p$ is essentially given by (\ref{large}).
\begin{figure}[ht]
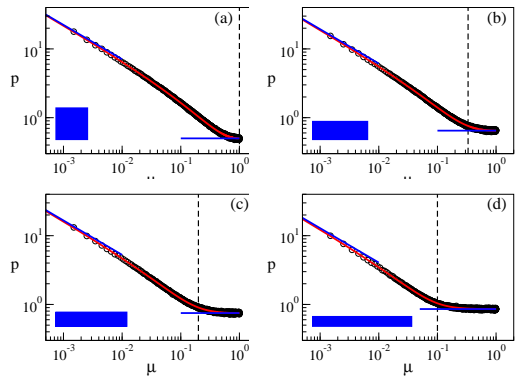

\includegraphics[width=1.3 in, angle=0]{fig4a.eps}
\includegraphics[width=1.3 in, angle=0]{fig4b.eps}
\includegraphics[width=1.3 in, angle=0]{fig4c.eps}
\includegraphics[width=1.3 in, angle=0]{fig4d.eps}
\caption{(color online) Log-log plots for $p(\mu)$.
The continuous curves refer to Eq. (\ref{density}) and
the open circles to numerical simulations.
The power-law regimes are evident in the four panels concerning the depicted rectangles 
($\gamma = 1, 3, 5, 10$).}
\label{fig4}
\end{figure}
All these features are shown in the log-log plots of Fig. \ref{fig4},
where the density $p$ is plotted against $\mu$ for aspect ratios of 1 (a), 3 (b), 5 (c), and 10 (d).
The continuous curve represents the analytical result of Eq. (\ref{density}) and
the open circles refer to numerical samplings with $5 \times 10^{6}$ minimal fragmentation events.
The dust regime is shown to behave according to the power law given by Eq. (\ref{dust}),
while for large fragments a flat region, Eq. (\ref{large}), is observed for $\mu > 1/\gamma$ 
(dashed vertical lines). 

\begin{figure}[ht]
\includegraphics[width=1.3 in, angle=0]{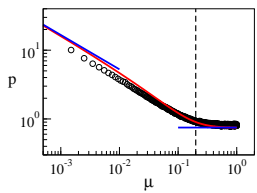}
\caption{(color online) Same as in Fig. \ref{fig4} for non-isotopic random segments. 
The continuous line represents Eq. (\ref{density}) without modification.}
\label{fig5}
\end{figure}

To verify whether these exponents are robust with respect to variations in the model, especially,
those which would make it more realistic, let us consider two relevant points. First, 
the triangle in Fig. \ref{fig1}(a) is smaller than the actual fragment. 
Assuming that the contrary situation is equally likely, we repeated our numerical calculations setting the new
mass as $m_{new}=m(1+\delta)$, where $\delta$ is a uniform random variable over a symmetric
interval $[-\epsilon,\epsilon]$. The results (not shown) are indistinguishable from those of the original model.
Second, although it has been often used in the literature, the isotropy condition may not
be strictly followed, for example, the crack may have a higher probability to propagate 
in the vicinity of the dashed line in Fig. \ref{fig1}(c). We then assumed that $\phi$
obeys a parabolic probability distribution, given by $\beta - \kappa \phi^2$ for 
$\phi \in [-\pi/2,\pi/2]$, with the normalization condition $\pi \beta = 1 + \pi^3 \kappa/12$, and
found that the exponents $1/2$ and $0$ are robust, the crossover region being shorter. 
In Fig. \ref{fig5} $\beta$ and $\kappa$ are such that the 
probability to find $\phi$ in the interval $[\pi/2, \pi/2-{\rm d}\phi]$ is $50\%$ smaller than that of
finding $\phi$ within $[0,{\rm d}\phi]$. 

In contrast, alternative definitions of the random variables themselves lead to very distinct results, showing that
our choice of a point on the perimeter is an essential physical assumption. We made calculations with the choice of 
an interior point plus an angle and the results do not resemble any data on fragmentation. In particular no power-law 
divergence has been observed. 

In conclusion, we addressed a new class of fragmentation problems. 
At first, it might be thought that the typical power-law divergence for small masses would not show up. 
Contrary to this
expectation, we obtained an exact exponent of $\alpha_1=0.5$ in the dust regime. A second power law
has been observed for large fragments with $\alpha_2=0$. These exponents are insensitive to the aspect ratio 
of the rectangles, and robust with respect to variations in the probability distributions used to generate the fragments.
In addition, we determined the shape abundances of the fragments for arbitrary aspect ratios. 
It is expected that this kind of minimal approach may stimulate new experimental realizations and theoretical efforts.
The presented model do not intend to describe all quantitative features of multifragmentation, even because the exponents are scattered over
a large range of experimental values. However, the simple model presented here describes experimental facts
like the absence of a characteristic size in the small mass regime and transitions due to effective dimensionality.
To better describe the range of exponents of multifragmentation, one possibility is to consider hierarchical models based on 
generations of minimal segmentation events, the number of generations being related to the energy input.

\begin{acknowledgments}
The authors thank the comments by M. Copelli, F. Brito, and M. Gomes. F. Santos and R. Donangelo
are gratefully acknowledged for the extra material kindly supplied regarding ref. \cite{santos}.
Funding from CNPq and FACEPE (APQ-1415-1.05/10) is gratefully acknowledged.
\end{acknowledgments}

\end{document}